\documentclass[aps,prb,onecolumn,showpacs,showkeys]{revtex4}
\usepackage{amssymb}
\usepackage{graphicx,color}
\usepackage{epsfig}
\usepackage{rotating}
\usepackage{amsmath}

\begin{document}

\title{A study of the influence of the mobility on the phase transitions of
the synchronous SIR model}
\author{Roberto da Silva$^{1}$, Henrique A. Fernandes$^{2}$}
\affiliation{$^{1}$Instituto de F{\'i}sica, Universidade Federal do Rio Grande do Sul,
Av. Bento Gon{\c{c}}alves, 9500 - CEP 91501-970, Porto Alegre, Rio Grande do
Sul, Brazil\\
$^{2}$Coordena{\c{c}}{\~a}o de F{\'i}sica, Universidade Federal de Goi{\'a}%
s, Campus Jata{\'i}, BR 364, km 192, 3800 - CEP 75801-615, Jata{\'i}, Goi{\'a%
}s, Brazil}

\begin{abstract}
By using an appropriate version of the synchronous SIR model, we studied the
effects of dilution and mobility on the critical immunization rate. We
showed that, by applying time-dependent Monte Carlo (MC) simulations at
criticality, and taking into account the optimization of the power law for
the density of infected individuals, the critical immunization necessary to
block the epidemic in two-dimensional lattices decreases as dilution
increases with a logarithmic dependence. On the other hand, the mobility
minimizes such effects and the critical immunizations is greater when the
probability of movement of the individuals increases.
\end{abstract}

\maketitle

\setlength{\baselineskip}{0.7cm}

\section{Introduction}

The study of critical phenomena in epidemic models defined on lattices has
an important hole in the context of Statistical Mechanics \cite%
{Grassberger1983, Dickman1999, Hinrichsen2000}. However, such approach is
relatively recent since the mathematical modelling of an epidemic process
was introduced via differential equations \cite{KermackMcKendrick}. The
advantages of the first approach in relation to differential equations
framework are the possibility of studying fluctuations, whereas it takes
into account correlations, and clustering effects.

In this context, the non-equilibrium phase transitions of systems that
possess absorbing states have been vastly studied, as can be seeing in the
literature (see, for example, Refs. \cite{Dickman1999,Hinrichsen2000}), and
can be didactically separated in two kinds: with and without immunization.

Epidemic models without immunization belong generically to the Directed
Percolation (DP) universality class. In this case, the finite size scaling
near criticality can be described by the following general scaling relation: 
\begin{equation*}
\left\langle i(t)\right\rangle \sim t^{-\beta /\nu _{\parallel }}f(\Delta
t^{1/\nu _{\parallel }},L^{-d}t^{d/z},d_{0}t^{\beta /\nu _{\parallel
}+\theta })
\end{equation*}%
where $i(t)=\frac{1}{L^{d}}\sum_{j=1}^{L^{d}}\sigma _{j}\ $is the density of
infected individuals in a $d-$dimensional lattice at the time $t$ and $%
\left\langle \cdots \right\rangle $ means the average on different
evolutions of the system. The individuals $\sigma _{j}$ are arranged on the
sites $j$ of the lattice (where $j=1,\cdots ,L^{d}$, and $L$ is the linear
size of the lattice) and can be in two different states, infected (or
active) and susceptible (or inactive). When the individual is infected, one
considers $\sigma _{j}=1$ and otherwise $\sigma _{j}=0$. The exponents $%
\beta $, $\nu _{\parallel }$, and $\nu _{\bot }$ are static critical
exponents, while $z=\nu _{\parallel }/\nu _{\bot }$ and $\theta $ are the
dynamic ones. Here, $\Delta =p-p_{c}$ denotes the distance of a point $p$ to
the critical point, $p_{c}$, which governs the algebraic behaviors of the
two independent correlation lengths: the spatial one which behaves as $\xi
_{\bot }\sim $ $\Delta ^{-\nu _{\bot }}$ and temporal one, $\xi _{\parallel
}\sim $ $\Delta ^{-\nu _{\parallel }}$. Basically, $\xi _{\bot }$ must be
thought of as the average over many independent realizations of the cluster
diameter while $\xi _{\parallel }$ is the same average of the required time
to reach the absorbing state.

Representing this universality class, we can consider, for example, the
susceptible-infected-susceptible\textbf{\ }(SIS) model which corresponds to
the Domany-Kinzel cellular automata when considering the synchronous update
and to the contact process when one takes into account the asynchronous
update. The SIS model represents diseases for which the infection does not
confer immunity, i.e., the individual return to the susceptible class after
recovering from the infection. For this model\textbf{,} we expect a
crossover between two power-law behaviors when starting the simulation with
a small density of infected sites\textbf{,} $i(t=0)=i_{0}\approx 0$, at
criticality ($\Delta =0$):

\begin{equation*}
\left\langle i(t)\right\rangle =\left\{ 
\begin{array}{ll}
t^{\theta } & \text{if\ }t<t_{0} \\ 
&  \\ 
t^{-\beta /\nu _{\parallel}z} & \text{elsewhere}%
\end{array}%
\right.
\end{equation*}
where the crossover time depends on $t_{0}\sim i_{0}^{-(1/\beta
/\nu_{\parallel}+\theta )}$. This critical initial slip is very similar to
the behavior of magnetic systems when they are quenched from a high
temperature to the critical one \cite{Janssen1989}.

Such behaviors can be studied from MC simulations. The first one (when $%
t<t_{0}$) is obtained by putting only one infected individual in the center
of the lattice and all other sites are occupied by susceptible ones ($%
i_{0}=1/L^{d}$). The second behavior can be obtained by making $i_{0}=1$
(representing a fully infected lattice). However, in models with
immunization, the initial condition $i_{0}=1$ does not reproduce the second
power law. Actually, in this case, an exponential relaxation is expected.

In addition to the SIS model, a range of other models can be studied in
order to represent a desired epidemic. For instance, one can consider the
susceptible-infected (SI) model to study the human immunodeficiency virus
(HIV) whereas there is no recovery. However, when the disease is such that
the individual can recover and be immune to the disease, at least for some
time, one must consider a new class of recovered individuals, R. If, after
recovering from the disease, the individual becomes immune to reinfection,
the epidemic can be studied through the susceptible-infected-recovered (SIR)
model. Nevertheless, if the individual can be reinfected with the disease
after some time, one can consider the SIRS model. In addition, there are
other models which demand other classes such as exposed (E), hidden (H), and
maternally-derived immunity (M), in order to represent infectious diseases.

The first studies about lattice-based models for epidemic growth is due to
Grassberger \cite{Grassberger1983}, who considered a simple cellular
automata to show that exists a critical contamination value such that, below
this point, no infinite epidemic is possible, i.e., the epidemic does not
percolate. In his pioneering work about epidemic model with immunization,
Grassberger \cite{Grassberger1983} raised important questions about epidemic
process, and one of them, in particular, caught our attention:
\textquotedblleft ... A somewhat more subtle question is whether one can
allow also for mobility of individuals...\textquotedblright . Unfortunately,
no work has explored the phase transitions and the preservation of some
scaling relations for epidemic models with immunization despite this
excellent tip.

In this work, we elaborate a detailed study of the influence of the mobility
of individuals on the phase transitions of the SIR model. For that, we
separate our main contributions in three parts as follows:

\begin{enumerate}
\item Elaboration of a refinement procedure to determine the critical
parameters via time-dependent Monte Carlo Simulations for models with
absorbing states;

\item The influence of dilution on the finite size scaling behavior of the
synchronous SIR and the dependence of the critical parameters on the density
of individuals in the lattice;

\item How the mobility can affect the parameters at the transition point,
more precisely the critical immunization probability ($c$)?
\end{enumerate}

Our work is divided in sections as follows: In the next section, we present
the SIR model implemented as cellular automata and show how to add the
dilution and mobility in this model. Moreover, we show how to adapt a
refinement procedure, developed in Ref. \cite{SilvaPRE2012} (when studying
spin systems in the context of time dependent MC simulations), in order to
determine the critical parameters of models with absorbing states. In Sec. %
\ref{Section:Results} we present our main results. Finally, a brief
discussion of the results as well as some conclusions of our findings are
presented in Sec. \ref{Section:Conclusions}.

\section{Sir model with dilution and mobility and time-dependent simulations}

\label{Section:time-dependent-simulations}

The susceptible-infected-recovered (SIR) model \cite{KermackMcKendrick} is a
paradigmatic model in the theory of epidemics, and can be considered as a
good and simple model to mimic some infectious diseases including measles,
mumps, and rubella. Defining $S(t)$ as the number of susceptible individuals
at time $t$ in a population, $I(t)$ the number of infected ones, and $R(t)$
the number of recovered (immune) ones, a deterministic (mean-field) approach
of the traditional SIR model considers the following set of differential
equations: 
\begin{eqnarray*}
\frac{dS}{dt} &=&-\beta I(t)S(t) \\
\frac{dI}{dt} &=&\beta S(t)I(t)-\gamma \ I(t) \\
\frac{dR}{dt} &=&\gamma I(t)\mathbf{,}
\end{eqnarray*}%
where $\beta $ is the contact rate which takes into account the probability
of getting the disease in a contact between a susceptible and an infected
subject, and $\gamma $ is simply the recovering rate of infected
individuals. This originally non-linear problem was proposed by Kermack and
MacKendrik and no generic analytic solution is known. Naturally, we can note
that $\frac{d}{dt}(S+I+R)=0$ which means that $N=S+I+R$ is a constant of the
problem: the total number of individuals. More precisely, this version of
the SIR model does not suppose vital dynamics, i.e., an epidemic where a
single epidemic outbreak moves faster than the normal birth/death rates.

In order to take into account the correlation and cluster effects in an
epidemic spreading, Grassberger \cite{Grassberger1983} considered a
synchronous probabilistic (cellular automaton) version of the epidemic model
on a regular lattice. A particular and interesting version of the
synchronous SIR model is obtained when the rates $\beta $ and $\gamma $ are
given by the probabilities $b$, the infection probability, and $c$, the
recovered one. In addition, in this particular version $b+c=1$ as it should.

In this version, each site can be in the following states: occupied by a
susceptible, an infected or an immune individual. The stochastic dynamic is
governed by the following rules: The infection can occurs when a susceptible
individual, which occupy a given site of the square lattice, has at least
one neighbor occupied by an infected individual. This process occurs with
probability $b/4$ times the number of infected individuals in its
neighborhood. The recovering process occurs spontaneously with probability $c
$ when the site is occupied by an infected individual. Recently, Arashiro
and Tome \cite{Arashiro2007} have proposed a version of the SIR model as a
particular case of the prey-predator model implemented as cellular automata.
In that work, they showed that when keeping the immunization rate ($c$) plus
infection rate ($b$) exactly equal to 1 and considering every site of the
lattice occupied with a susceptible, infected or recovered individual, a
transition between the active and inactive phases occurs for $c_{c}=0.22$.
This is the critical value of the infection probability, where occurs the
transition from a phase where the density of recovered individuals, $r(t)$,
is equal to 1 (meaning that $b$ is high enough such that the whole
population, after contaminated, become recovered) to a phase where $r(t)=0$
(whereas $b$, in this case, is low enough such that only few individuals, or
even no one of them, become recovered). The authors performed several
independent runs of a lattice completely filled with susceptible individuals
except by one infected site at the center of the lattice and the system
evolved according to the synchronous update. After a reasonable number of
steps, any one of the infinitely many absorbing states can be reached and
consequently, the number of immune individuals at the steady state varies
from run to run. The simulation is finished when the system enters in an
absorbing state. Then, the number of recovered individuals is calculated.
The mean value of this quantity, divided by the total number of individuals,
is the density of immune individuals at the steady state, $\left\langle
r(t)\right\rangle $, as presented above.

Although this model has been considered for the case where the density of
individuals is $\rho =1$, i.e., no vacancies are presented in the lattice,
in many real applications some sites of the lattice can be empty, affecting
therefore the critical properties of the model. Previous studies in other
Statistical Mechanical models (see for example diluted Ising model \cite%
{Mazzeo1999,Plascak2007}) show that phase transitions are changed when $\rho
\neq 1$. By extending the applications for other diluted lattices, Vainstein 
\textit{et al}. \cite{Vainstein2007} explored not only the influence of
dilution on the cooperation among prisoner dilemma players but also the
effects of the mobility of players. Vainstein \textit{et al.} \cite%
{Vainstein2007} considered the mobility simulated by a simple random walk in
the two-dimensional lattice. In their prescription, each player can only
jump to its nearest neighbor at random and when there is at least one
nearest-neighbor site empty. Thus, this jump occurs with probability $p$.
Otherwise, the player remains where it is.

For an epidemic process, it is not different and infections must be changed
not only by dilution but mainly by mobility! In this paper, we analyse the
influence of mobility of individuals on the phase diagrams of the SIR model.
For this task, we consider its synchronous version implemented according to
Ref. \cite{Arashiro2007}. On the other hand, the mobility was implemented
according to the prescription used in Ref. \cite{Vainstein2007}.

For the refinement process of the critical parameters, we use a method
developed in Ref. \cite{SilvaPRE2012} and vastly explored in our previous
contributions in which we studied magnetic systems \cite%
{SilvaPRE2013,SilvaCPC2013,SilvaPRE2014}. To our knowledge, this is the
first time that this kind of refinement is applied to a model without a
defined Hamiltonian.

Basically, we perform time dependent simulations by starting with the system
completely filled with susceptible individuals but the individual at the
center of the lattice which is contaminated with the disease. By considering 
$N_{\text{runs}}$ different runs, we calculate the average time of the
density of infected individuals of a square lattice of linear size $L$
through the equation 
\begin{equation*}
\left\langle i(t)\right\rangle =\frac{1}{N_{run}}\sum_{k=1}^{N_{run}}\frac{%
\sum_{j=1}^{L}\sum_{l=1}^{L}m_{l,j,k}(t)\delta _{1,\sigma _{l,j,k}(t)}}{%
\sum_{j=1}^{L}\sum_{l=1}^{L}m_{l,j,k}(t)}\mathbf{.}
\end{equation*}%
Here, $m$ is 0 when the site $(l,j)$ is empty or 1 when it is occupied by an
individual at the time $t$ of the $k-$th run. In addition, $\sigma $ denotes
the state of an individual: susceptible ($\sigma =0$), infected ($\sigma =1$%
), or recovered ($\sigma =2$), $L$ is the linear size of the lattice and $%
l=1,\cdots ,L$ $(j=1,\cdots ,L)$ stands for the $l-$th line ($j-$th column)
of the lattice.

Since at criticality it is expected that $\left\langle i(t)\right\rangle
\sim t^{\theta }$, the method/algorithm changes the values of the
immunization rate, $c$, according to a resolution $\Delta c$ from $c_{\min }$
up to $c_{\max }$ in order to find its best value, i.e., the critical
immunization probability, $c_{c}$. Then, we calculate the known coefficient
of determination \cite{Trivedi2002} 
\begin{equation}
\alpha =\frac{\sum_{t=t_{\min }}^{N_{MC}}\left( \overline{\ln \left\langle
i\right\rangle }-a_{1}-a_{2}\ln t\right) ^{2}}{\sum_{t=t_{\min
}}^{N_{MC}}\left( \overline{\ln \left\langle i\right\rangle }-\ln
\left\langle i(t)\right\rangle \right) ^{2}}  \label{determination}
\end{equation}%
for each value of $c=c_{\min }+$ $i\ \Delta c$, where $i=1,...,\left\lfloor
(c_{\max }-c_{\min })/\Delta c\right\rfloor $ and $N_{MC}$ is the number of
MC steps. Here $a_{1}$ and $a_{2}$ come from the linear fit of $\ln
\left\langle i(t)\right\rangle $ versus $\ln t$, and $\overline{\ln
\left\langle i\right\rangle }$ $=(1/(N_{MC}-t_{\min }+1))\sum_{t=t_{\min
}}^{N_{MC}}\ln \left\langle i\right\rangle (t)$. In addition, $t_{\min }$ is
the number of disregarded MC steps at the beginning of the simulation. The
coefficient $\alpha $ has a very simple explanation: It measures the ratio
(expected variation)/(total variation). The bigger the $\alpha $, the better
the linear fit in log scale, and therefore, the better the power law which
corresponds to the critical parameter excepted for an error $O(\Delta c)$.

\section{Results}

\label{Section:Results}

\begin{figure}[th]
\begin{center}
\includegraphics[width=0.33\columnwidth]{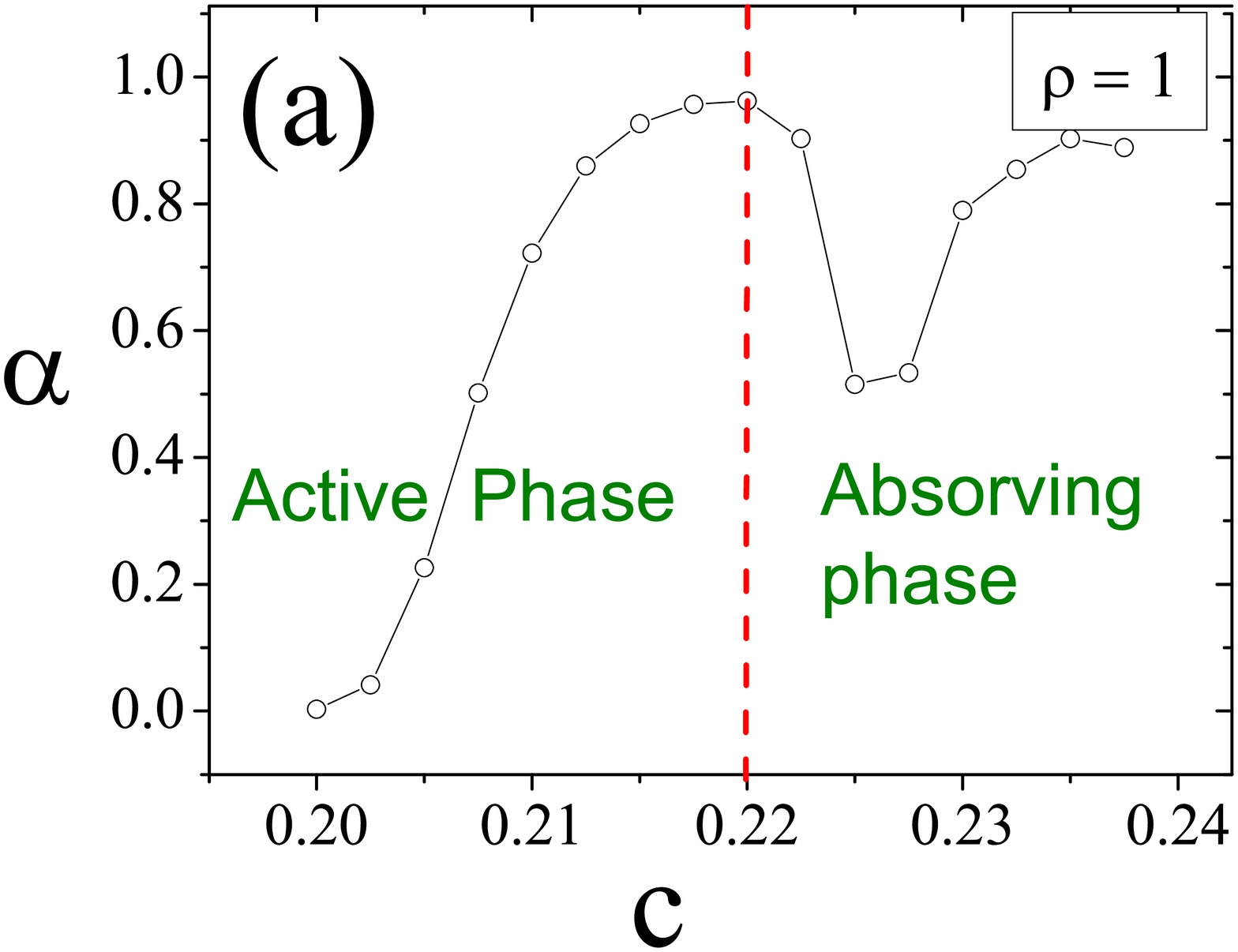}%
\includegraphics[width=0.33\columnwidth]{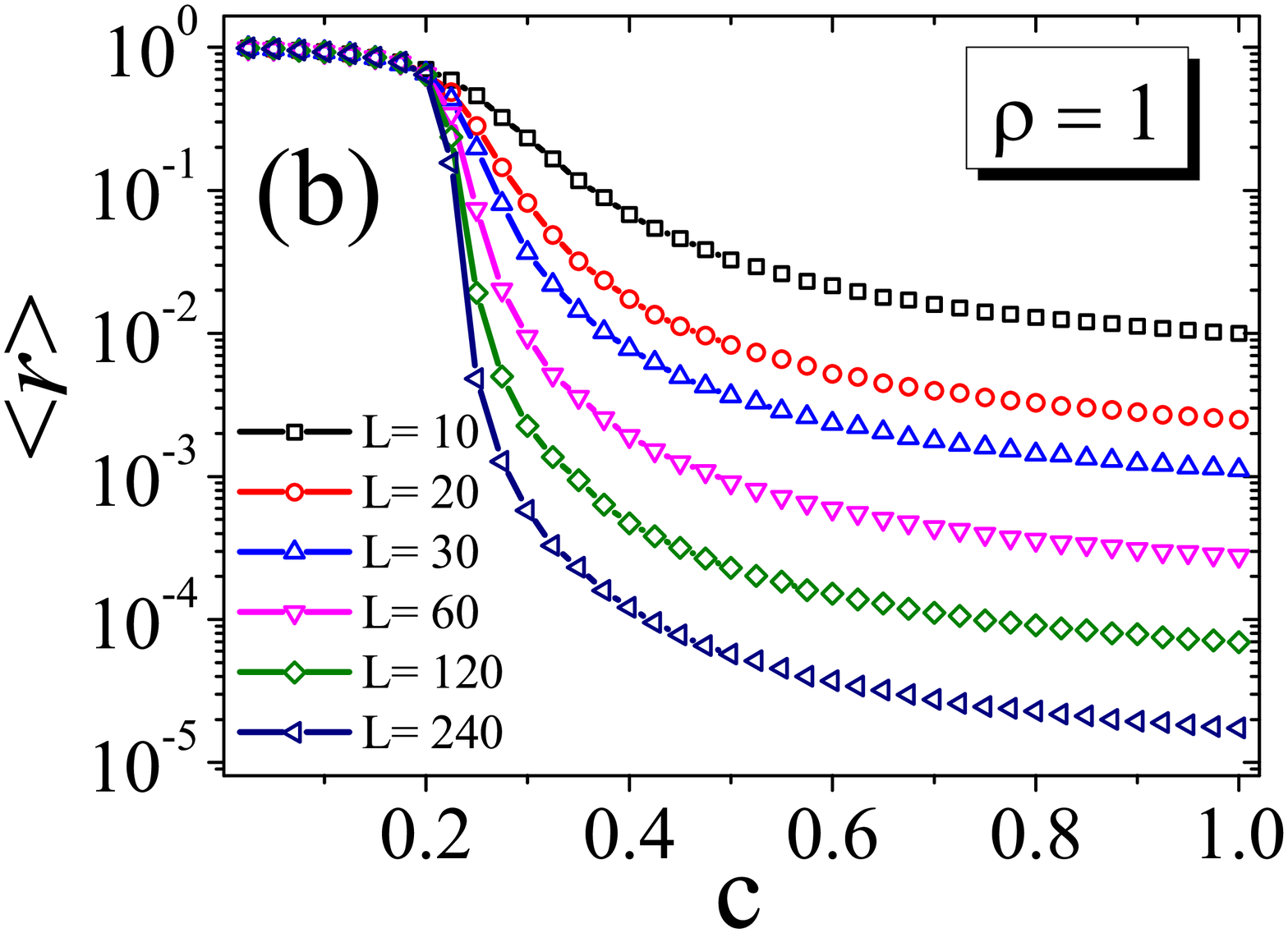}%
\includegraphics[width=0.33\columnwidth]{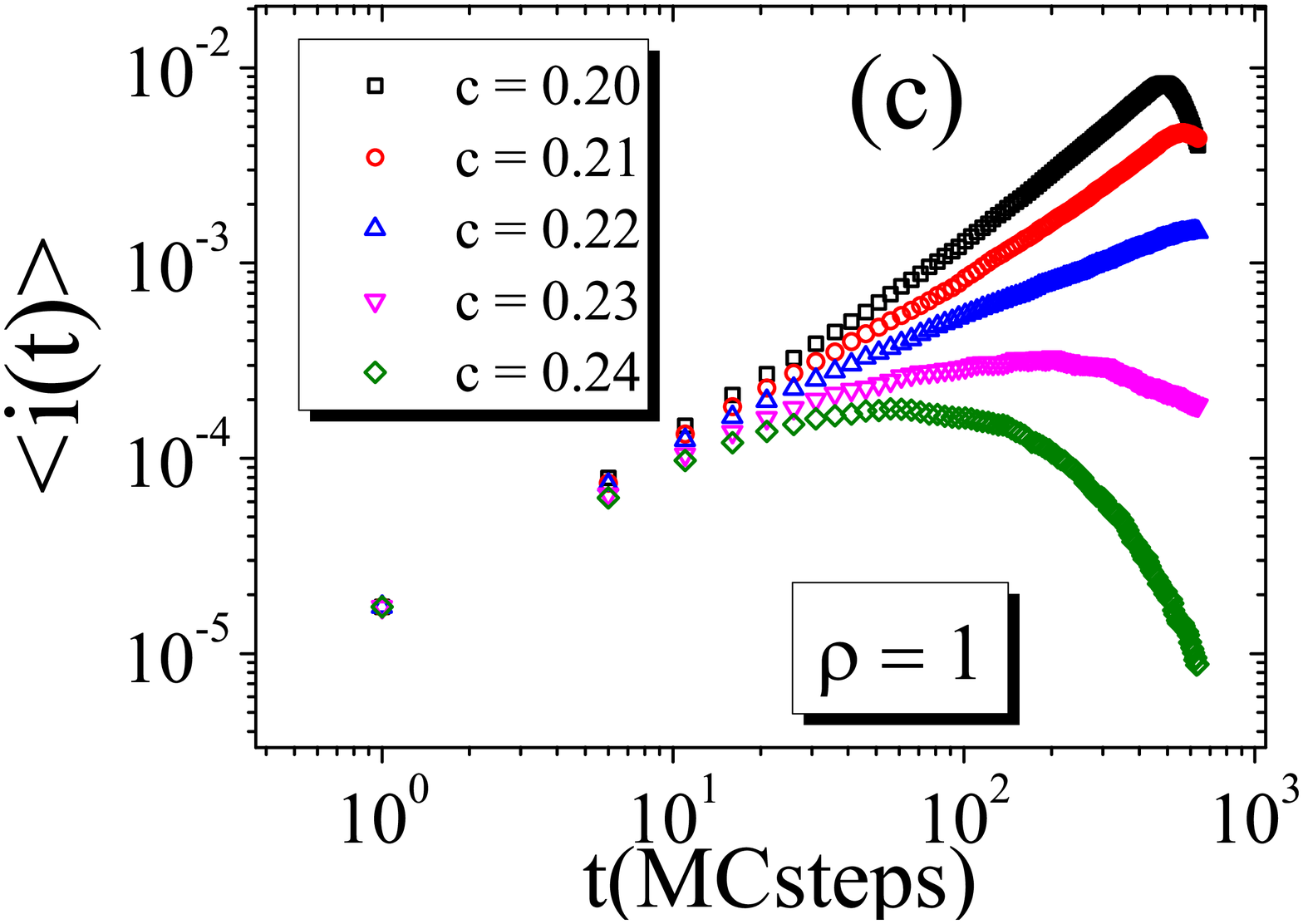}
\end{center}
\caption{\textbf{Figure (a)}: Values of the coefficient of determination $%
\protect\alpha $ as function of $c$ for the case $\protect\rho =1$. \textbf{%
Figure (b)}: Finite size scaling for the density of recovered individuals as
function of $c$. The phase transition is observed in $c=0.22$. \textbf{%
Figure (c)}: The search for a power law behavior of $\left\langle
i(t)\right\rangle $ for different values of $c$ around $c_{c}=0.22$ via
time-dependent Monte Carlo simulations. The value of $c$ which corresponds
to the curve with linear behavior corroborates the value found in Figs. (a)
and (b).}
\label{phase_diagram_ro_eq_0}
\end{figure}

By using the optimization method presented in the previous section, we
determined the critical immunization probability ($c_{c}$) for different
values of dilution occupation $0\leq \rho \leq 1$ and mobility $0\leq p\leq
1 $. For time-dependent simulations, we used $t_{\min }=100$ MC steps and $%
N_{MC}=800$ MC steps. Let us start with the simplest case where $\rho =1$
which corresponds to the critical value $c_{c}=0.22$ obtained by Arashiro
and Tom\'{e} \cite{Arashiro2007} whereas there is no empty site in the
lattice. As shown in Fig. \ref{phase_diagram_ro_eq_0} (a), our simulations
pointed out for a maximum of $\alpha $ corresponding to $c_{c}=0.2200(25)$
which corroborates the value from literature. Basically, our refinement
process changes $c$ with a precision $\Delta c=0.0025$ and in addition, the
time-dependent simulations are performed with $N_{run}=2400$ runs for each
value of $c$.

Fig. \ref{phase_diagram_ro_eq_0} (b) shows that the model presents a phase
transition for $c_{c}=0.22$ corroborating the value found in Fig. \ref%
{phase_diagram_ro_eq_0} (a). This is shown by calculating $\left\langle
r\right\rangle $, the average of recovered individuals, in the steady state
(absorbing state). We considered for these simulations $N_{run}=4000$ and
the maximum number of MC steps $N_{MC}^{(\max )}=3000$ to ensure that the
steady state is reached for each value of $c$ studied. Such a result
corroborates the result obtained by the refinement process (Fig. \ref%
{phase_diagram_ro_eq_0} (a)). The optimization described by Fig. \ref%
{phase_diagram_ro_eq_0} (a) can be visually observed in Fig. \ref%
{phase_diagram_ro_eq_0} (c).

Since we observed that the optimization method described in Sec. \ref%
{Section:time-dependent-simulations} works for the standard case, we decided
to apply this approach to study the phase transitions of the SIR model when
there is dilution of the lattice and mobility of the individuals. First of
all, it is worth to describe how the dilution effects can change the
critical immunization rates. In order to study the dependence of $c_{c}$
with $\rho $ when $p=0$ (static scenarios), we changed $\rho $ from $1$ up
to $0.70$ by considering $\Delta \rho =0.05$. We applied the refinement
process and determined the best $c$ (corresponding to the highest $\alpha $)
for each $\rho -$value.

\begin{figure}[th]
\begin{center}
\includegraphics[width=0.5\columnwidth]{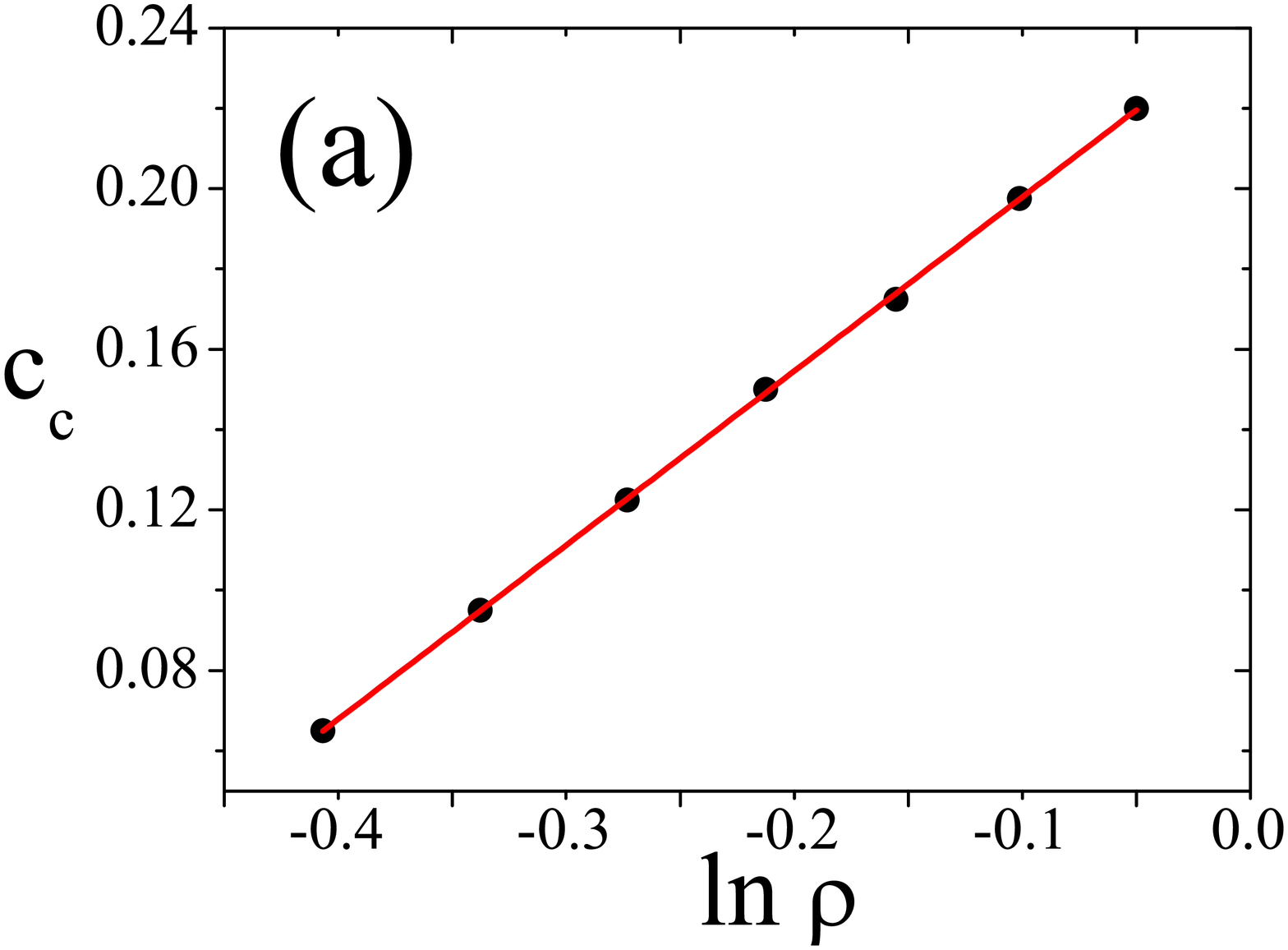}%
\includegraphics[width=0.5\columnwidth]{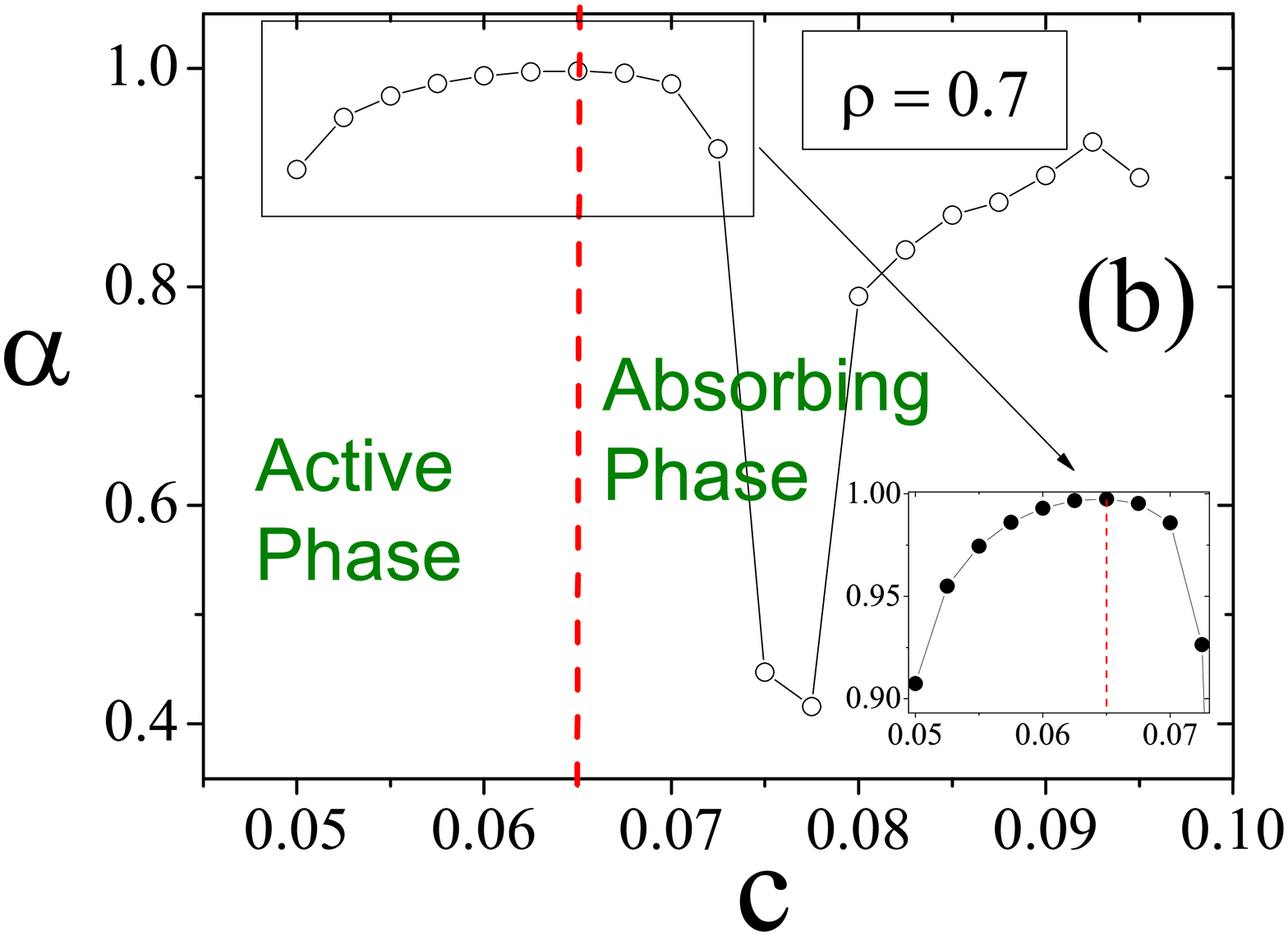}
\end{center}
\caption{\textbf{Figure (a)}: Critical Immunization rate ($c_{c}$) versus
the density of individuals ($\protect\rho $) for $p=0$. We can observe a
linear dependence on $\ln \protect\rho $. \textbf{Figure (b)}: Illustration
of one case of the refinement process for $p=0$ and $\protect\rho =0.7$.}
\label{diluited_no_mobile}
\end{figure}

Fig. \ref{diluited_no_mobile} (a) shows the behavior of $c$ as function of $%
\ln \rho $. We observed a logarithm law for $p=0$ where $c_{c}=a+b\ln \rho $%
, with $a=0.2196(5)$ and $b=0.434(2)$ which produces a very good fit: $%
\alpha =0.9998$. The plot (b) of Fig. \ref{diluited_no_mobile} shows the
optimization curve ($\alpha \times c$) for the more diluted case ($\rho =0.7$%
). Since we have already studied the case $p=0$ (without mobility), now we
turned our attention to the mobility of the individuals in the lattice. By
using the same procedure for each value of $\rho $, we performed simulations
in order to obtain the critical immunization $c$ as function of $p$.
Firstly, we can observe in Fig. \ref{rhoeq0,85_and_peq0,3} (a) the phase
diagram for the case where $\rho =0.85$ and $p=0.30$. 
\begin{figure}[th]
\begin{center}
\includegraphics[width=0.5\columnwidth]{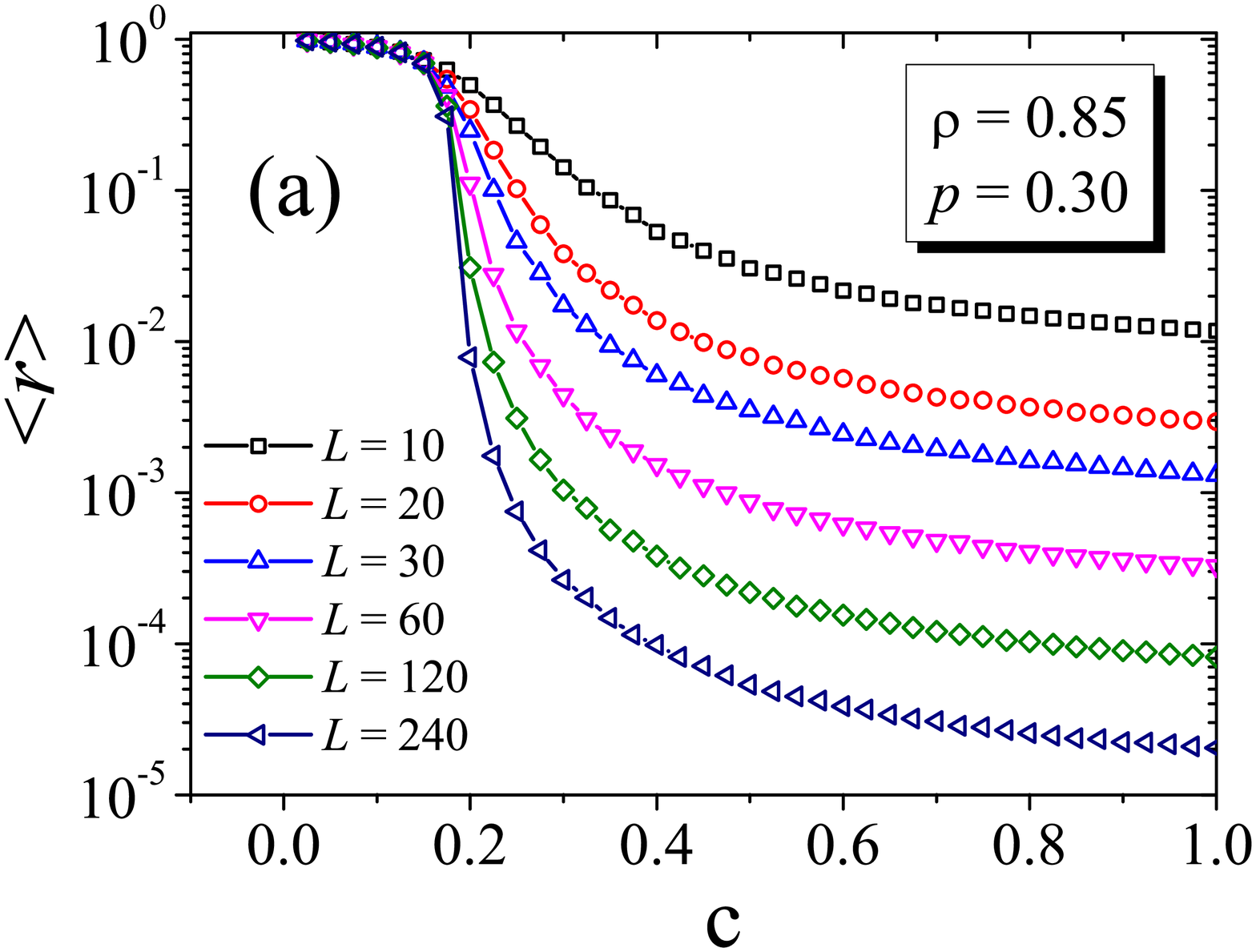}%
\includegraphics[width=0.5\columnwidth]{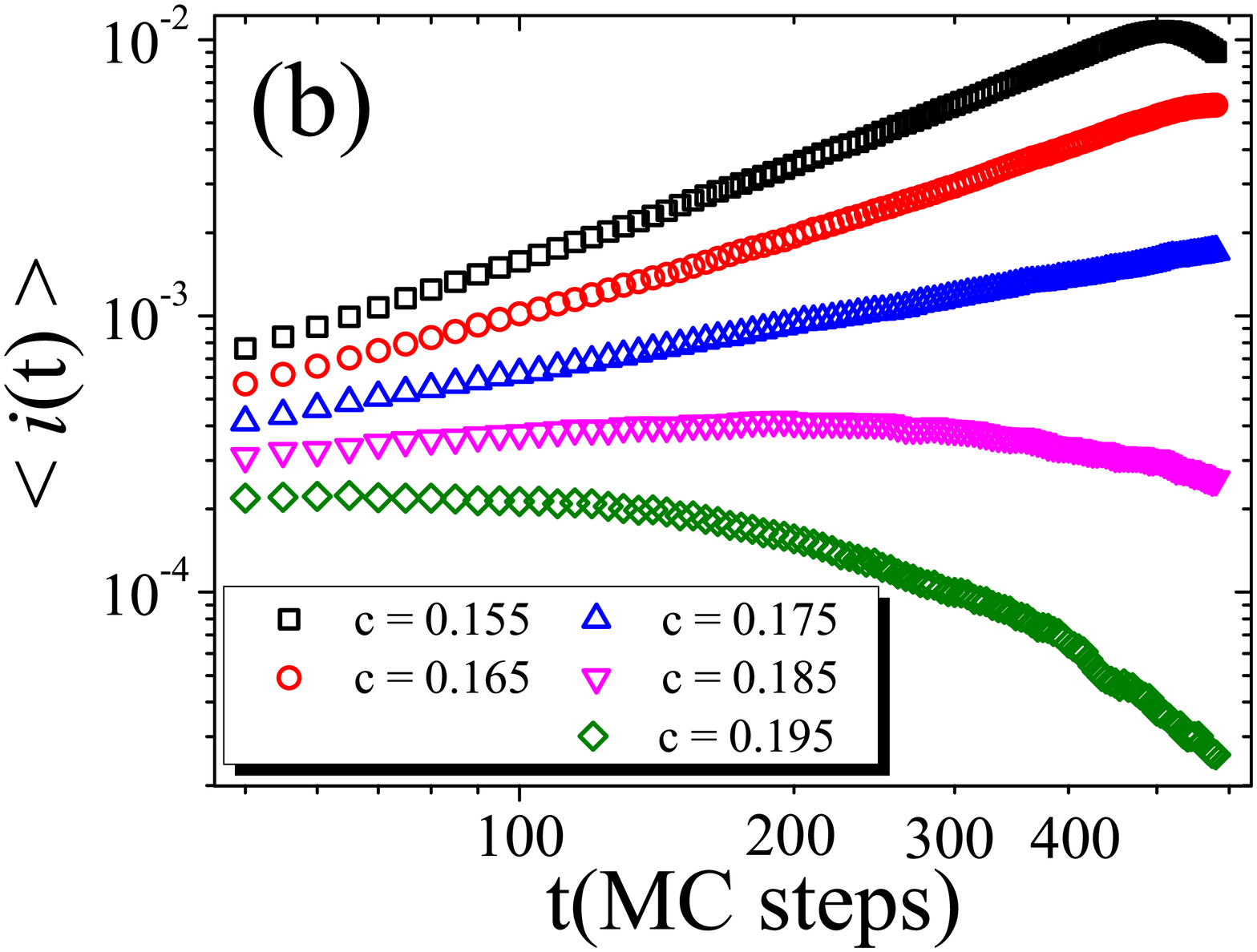}
\end{center}
\caption{\textbf{Figure (a)}: Phase diagram for $\protect\rho =0.85$ and $%
p=0.3$. \textbf{Figure (b)}: Illustration of the time-dependent simulations
for $\protect\rho =0.85$ and $p=0.3$.}
\label{rhoeq0,85_and_peq0,3}
\end{figure}

\begin{figure}[th]
\begin{center}
\includegraphics[width=0.7\columnwidth]{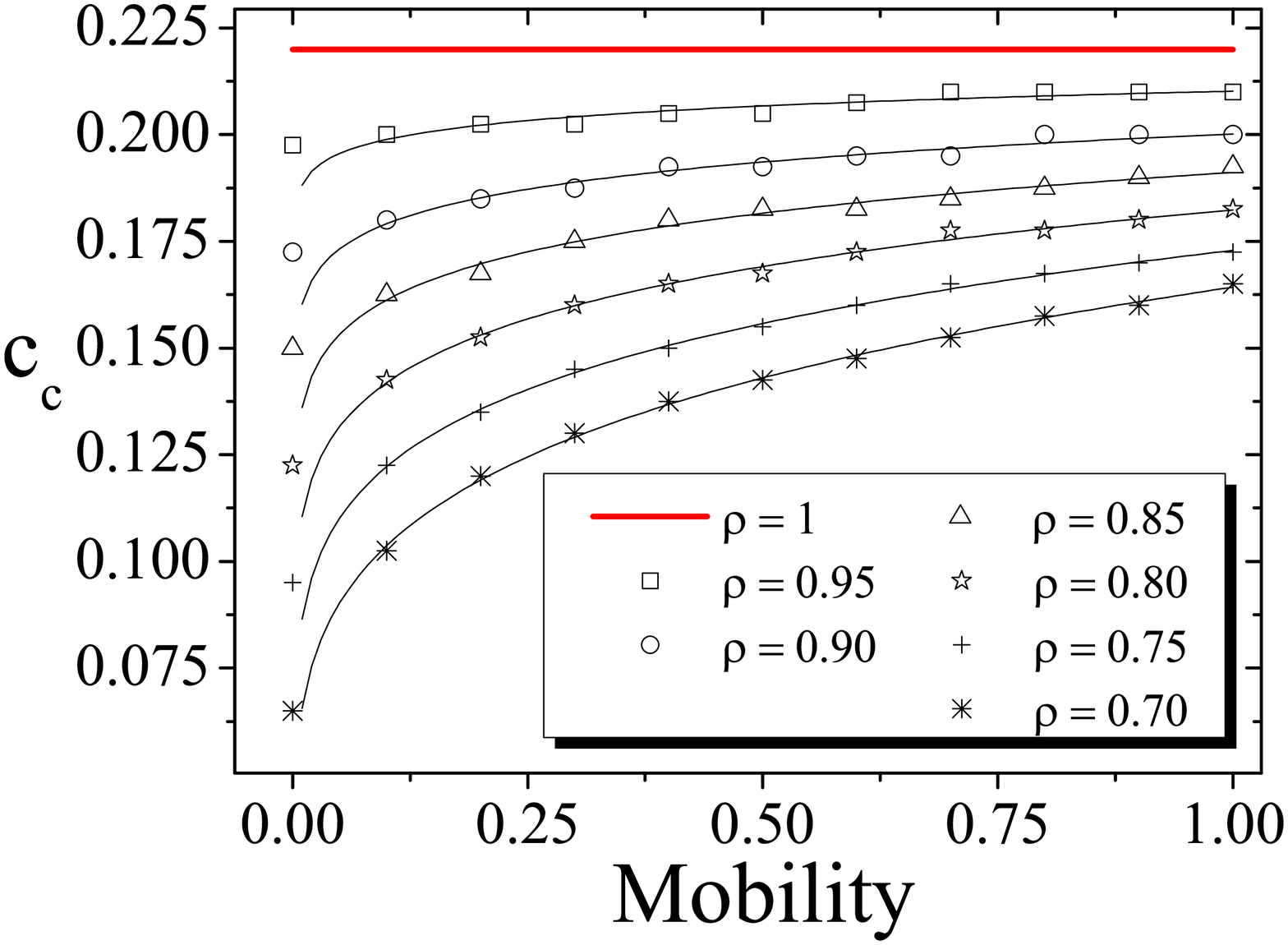}
\end{center}
\caption{Study of mobility effects on the critical immunization rates. Good
fits with power functions $c(\protect\rho )=\protect\gamma (\protect\rho )p^{%
\protect\beta (\protect\rho )}$ are observed.}
\label{mobility_effects}
\end{figure}

The phase transition occurs for $c_{c}=0.1750(25)$ which is corroborated by
time-dependent simulations as can be observed in the plot (b) of Fig. \ref%
{rhoeq0,85_and_peq0,3}. According to our previous study, when $p=0$, and for
the same value of $\rho $ ($\rho =085$), we obtained $c_{c}=0.1500(25)$.
Hence, it suggests that the critical immunization rates must depend of $p$
for fixed values of $\rho $. We explore such a dependence as can be seen in
Fig. \ref{mobility_effects}.

In this figure, we can see that $c$ increases as $p$ increases. For $\rho $
fixed, $c_{c}$ was empirically fitted by a power function $c(\rho )=\gamma
(\rho )p^{\beta (\rho )}$. The values $\gamma (\rho )$ and $\beta (\rho )$
are reported in Table \ref{Table:exponents}.

\begin{table}[tbp] \centering%
\begin{tabular}{llllllll}
\hline\hline
$\rho $ & $1.0$ & $0.95$ & $0.90$ & $0.85$ & $0.80$ & $0.75$ & $0.70$ \\ 
\hline\hline
$\gamma $ & $\mathbf{0.22}$ & $0.210(1)$ & $0.200(1)$ & $0.191(1)$ & $%
0.182(1)$ & $0.172(1)$ & $0.164(1)$ \\ 
$\beta $ & $\mathbf{0.00}$ & $0.024(3)$ & $0.048(3)$ & $0.074(4)$ & $0.109(4)
$ & $0.151(2)$ & $0.200(1)$ \\ \hline\hline
\end{tabular}%
\caption{Exponents obtained from curves of critical immunization $c_{c}$ as
function of $p$ for different density of individuals}\label{Table:exponents}%
\end{table}%

So, we can assert that critical immunizations increases as $p$ increases.
However, for $\rho \neq 1$, $c_{c}$ never is greater than $0.22$ which
corresponds to $\rho =1$, even for maximum mobility ($p=1$).

\section{Summaries and conclusions}

\label{Section:Conclusions}

Based on two-dimensional SIR model with synchronous update, we performed (by
following a suggestion of the seminal paper of Grassberger \cite%
{Grassberger1983}) a detailed study of dilution and mobility effects on the
critical immunizations $c$, i.e., values of the recovering probability for
which the epidemic transits from a not controlled regime (where all
individuals are contaminated at sometime) to a controlled regime (few
individuals or none at all get sick). We used a method to refine the
critical parameters based on the optimization of the power law for the
density of infected individuals along the time at criticality. Our work
shows that the dilution of the lattice decreases the immunization $c$ based
on a log-law. However, this effects is minimized when the individuals have
the possibility to move on the lattice. In this case, for a fixed density of
individuals, the greater the mobility, the bigger the immunization rate.

\textbf{Acknowledgments --} This research was partially supported by the
Conselho Nacional de Desenvolvimento Cient\'{\i}fico e Tecnol\'{o}gico
(CNPq), under the grant 11862/2012-8. The authors would like to thank Prof.
L.G. Brunet (IF-UFRGS) for kindly providing the computational resources from
Clustered Computing (ada.if.ufrgs.br) for this work.

\end{document}